\documentstyle[preprint,aps]{revtex}
\begin{document}
\begin{center}
{\bf R. Beck and H.-P. Krahn Reply: $E2/M1$ ratio from the Mainz
$p(\vec{\gamma},p)\pi^0$}
\end{center}

In a recent Letter \cite{beck}
we have reported precision measurements
of differential cross sections and polarized photon asymmetries for the
reaction $\vec{\gamma}p$ $\rightarrow$ $p\pi^0$ with the DAPHNE--detector, 
using tagged photons at the Mainz Microtron MAMI.
The above Comment of Workman \cite{workman} raises 3 points which we will
discuss in turn:

(i) It correctly notes a typographical error
( $| E_{0+} |^2 + | 3E_{1+} + M_{1^+} - M_{1^-} |^2$ instead of
$| E_{0+} |^2 + |3E_{1+} - M_{1^+} + M_{1^-} |^2$ )
in our Eq.~(4) which, however, has not been made in the analysis and 
consequently has no effect on our result for $R_{EM}$.

(ii) It critizies the systematic error for $R_{EM}$.
We are using Eqs.~(3) to (7) in our paper \cite{beck} to extract the $E2/M1$
ratio. These Eqs. are exact under the assumption that only s-- and p--waves
contribute. 
The inclusion of d--waves results in a modification of Eq.~(3) to
\begin{eqnarray}
\frac{d\sigma}{d\Omega} = \frac{q}{k}( A + B cos(\theta) + C cos^2(\theta) +
D cos^3(\theta) + E cos^4(\theta)) ~~. \; 
\end{eqnarray}
Two additional coefficients D and E appear and furthermore the
coefficients A, B and C are modified according to
\begin{eqnarray}
A &\simeq& A(s_{wave},p_{wave}) +  {\rm Re} \left[E_{0+}d^*\!\!_{wave}  \right] +  
| d_{wave} |^2  \; , \\
B &\simeq& B(s_{wave},p_{wave}) + {\rm Re} \left[ (M_{1+}-M_{1-}) d^*_{wave}  \right] \; , \\
C &\simeq& C(s_{wave},p_{wave}) + {\rm Re}\left[ E_{0+} d^*\!\!_{wave} \right] + |
d_{wave} |^2 \; , \\
D &\simeq& {\rm Re} \left[ (M_{1+}-M_{1-}) d^*\!\!_{wave} \right] \; , \\
E &=& | d_{wave} |^2 \; , 
\end{eqnarray}
where $s_{wave},~p_{wave}$ and $d_{wave}$ are combinations of the 
corresponding partial wave multipoles. 
The effect is largest for the coefficients B and D, where an interference term
between the large $M_{1+}$ and the d--waves occurs. But at the top of the
resonance the contributions of these terms can be neglected, e.g.
\begin{eqnarray}
{\rm Re} \left[ (M_{1+}-M_{1-}) E_{2-}^* \right] &=& 
{\rm Re} (M_{1+}-M_{1-}) {\rm Re} E_{2-} + 
{\rm Im} (M_{1+}-M_{1-}) {\rm Im} E_{2-} ~~~.  \;
\end{eqnarray}
The first term vanishes, because ${\rm Re} (M_{1+}-M_{1-})$ goes through zero
near the resonance energy ($E_{\gamma} = 340$ MeV) and the second term can be
neglected, because ${\rm Im} E_{2-}$ is small due to a phase close to zero.
Below and above the resonance, contributions from $L \ge 2$ are of the order
of $10 - 20 \%$ of the differential cross section at $0^0$ and $180^0$.
This will affect the $C_{\|}$--coefficient below and above the resonance
energy.

The second reason, why $E_{\gamma} = 340$ MeV is special, is that at this point
one only has contributions from isospin $3/2$ to the ratio $R = C_{\|} /
12A_{\|}$ for the ($p,\pi^0$)--channel and thus 
\begin{eqnarray}
R \simeq \frac{ ImE^{3/2} \!\!\!\!\!\!\! _{1+} } { ImM^{3/2}
\!\!\!\!\!\!\! _{1+} } = R_{EM} ~~~.\; 
\end{eqnarray}
In the Comment \cite{workman} it is suggested, that our systematic error
($\pm 0.2 \%$) for $R_{EM}$ should be enlarged because neglecting 
all contributions apart those involving $| M_{1+} |^2$  and ${\rm
Re}(M_{1+}E_{1+}^*)$ in $C_{\|}$ and $A_{\|}$
results in
\begin{eqnarray}
R = \frac{C_{\|}} {12A_{\|}} \simeq \frac{R_{\pi^0}}{1 - 6R_{\pi^0}} ~\rightarrow ~
R_{\pi^0} = -2.95\% ~~~, \; 
\end{eqnarray}
with
\begin{eqnarray}
R_{\pi^0} = \frac{ImE_{1^+}} {ImM_{1^+} - ImM_{1^-}} ~~~. \; 
\end{eqnarray}
However neglecting
$R_{\pi^0}$ in the denominator is reduced by a cancellation between ${\rm
Im}M_{1-}$ and ${\rm Im}E_{1+}$ in $A_{\|}$ and the isospin
$1/2$ contribution of
${\rm Im}E_{1+}$ in $C_{\|}$. As a result, one obtains at $E_\gamma = 340$ MeV
($\delta_{33} = 90^0$)
\begin{eqnarray}
R = \frac{C_{\|}} {12A_{\|}} \simeq \frac{1.1R_{EM}}{1 - 6.6R_{EM}} 
~\rightarrow~ R_{EM} = -2.65\% ~~~,\; 
\end{eqnarray}
which is well within our quoted systematic error.

(iii) We believe that the reason for the different $R_{EM}$ value 
-($1.5 \pm0 .5) \%$ in the VPI multipole analysis program SAID is partly 
due to the used database. This is supported by a recent fixed--t dispersion 
relation multipole analysis by Hanstein {\it et al.} \cite{Hanstein}
based on the new ELSA, MAMI and recent TRIUMF data for
$\pi^0$, $\pi^+$ and $\pi^-$ production on the nucleon.
Their value for $R_{EM} = -2.4 \% $ is in good agreement with 
our value $R_{EM} = -(2.5 \pm0.2 \pm 0.2)\%$
(see also \cite{arndt}).

R. Beck and H.-P. Krahn \\
{\it
 Institut f\"ur Kernphysik der Universit\"at Mainz,\\
 Becherweg 45, 55099 Mainz, Germany}

{\large{\bf Appendix}}

At $E_{\gamma} = 340$ MeV one has $ReM_{1^+}(3/2) = 0$, $\delta_{33} = 90^0$, 
$Re(M_{1^+} - M_{1^-}) \simeq 0$ and contributions from higher partial waves
($L \ge 2$) can be neglected (see point ii).
\begin{eqnarray}
R = \frac{C_{\|}} {12A_{\|}} = \frac{Re(E_{1^+}(M_{1^+} - M_{1^-})^*)}
{|E_{0^+}|^2 + |3E_{1^+} - M_{1^+} + M_{1^-}|^2} ~~~. \; 
\end{eqnarray}
Neglect $|E_{0^+}|^2$, $9|E_{1^+}|^2$ and all terms with $Re(M_{1^+} -
M_{1^-})$ results in:
\begin{eqnarray}
R = \frac{ImE_{1^+}Im(M_{1^+} - M_{1^-})}
{-6ImE_{1^+}Im(M_{1^+} - M_{1^-}) + Im^2(M_{1^+} - M_{1^-})} ~~~. \; 
\end{eqnarray}
Divide by $Im^2(M_{1^+} -M_{1^-})$ gives
\begin{eqnarray}
R = \frac{R_{\pi^0}}{1 - 6R_{\pi^0}} ~\rightarrow ~
R_{\pi^0} = -2.95\% ~~~ for ~~~ R = -2.5\% , \; 
\end{eqnarray}
with
\begin{eqnarray}
R_{\pi^0} = \frac{ImE_{1^+}} {ImM_{1^+} - ImM_{1^-}} ~~~. \; 
\end{eqnarray}
Neglect $ImM_{1^-}$, and the 1/2 isospin components $ImM_{1^+}(1/2)$ and 
$ImE_{1^+}(1/2)$.
The largest correction comes from $ImE_{1^+}(1/2)$, which is of the order of
$10-20\%$ of $ImE_{1^+}$.
The final result is:
\begin{eqnarray}
R = \frac{C_{\|}} {12A_{\|}} \simeq \frac{1.1R_{EM}}{1 - 6.6R_{EM}} 
~\rightarrow~ R_{EM} = -2.65\% ~~~ for ~~~ R = -2.5\% , \; 
\end{eqnarray}
with
\begin{eqnarray}
R_{EM} = \frac{ ImE^{3/2} \!\!\!\!\!\!\! _{1+} } { ImM^{3/2}
\!\!\!\!\!\!\! _{1+} } ~~~.\; 
\end{eqnarray}

\end{document}